\documentclass{aa}

\usepackage{rotating}
\usepackage{graphicx}
\usepackage{txfonts}
\usepackage{natbib}
\usepackage[usenames]{color}

\defcitealias{goobar08a}{Paper II}

\begin{document}
\title{Near-IR search for lensed supernovae behind galaxy clusters }
\subtitle{I. Observations and transient detection efficiency}

\author{V.~Stanishev\inst{1}, A.~Goobar\inst{2,3},  K.~Paech\inst{2,3}, 
   R.~Amanullah\inst{2,3},T.~Dahl\'en\inst{4},
   J.~J\"onsson\inst{5}, J.~P.~Kneib\inst{6},
   C.~Lidman\inst{7}, M.~Limousin\inst{6,8}, E.~M\"ortsell\inst{2,3},
   S.~Nobili\inst{2}, J.~Richard\inst{9}, T.~Riehm\inst{10,3}, \and M.von
   Strauss\inst{2,3} \fnmsep\thanks{Based on observations made
          with ESO telescopes at the Paranal Observatory
          under program IDs 079.A-0192 and 081.A-0734.}}

\institute{
CENTRA - Centro Multidisciplinar de Astrof\'isica, Instituto Superior T\'ecnico, Av.
Rovisco Pais 1, 1049-001 Lisbon, Portugal
\and
Department of Physics, Stockholm University, Albanova University Center, S--106 91 Stockholm, Sweden 
\and
The Oskar Klein Center, Stockholm University, S--106 91 Stockholm, Sweden 
\and
Space Telescope Science Institute, Baltimore, MD 21218, USA
\and 
University of Oxford Astrophysics, Denys Wilkinson Building, Keble Road, Oxford OX1 3RH, UK
\and
Laboratoire d'Astrophysique de Marseille, OAMP, CNRS-Universit\'e Aix-Marseille,
38, rue Fr\'ed\'eric Joliot-Curie, 13388 Marseille cedex 13, France
\and
ESO, Vitacura, Alonso de Cordova, 3107, Casilla 19001, Santiago, Chile
\and 
Dark Cosmology Centre, Niels Bohr Institute, University of Copenhagen, Juliane Maries Vej 30, DK-2100 Copenhagen, Denmark
\and 
Institute for Computational Cosmology, Department of Physics and Astronomy, University of Durham,
    South Road, Durham, DH1 3LE, UK
\and 
Department of Astronomy, Stockholm University, Albanova
University Center, S--106 91 Stockholm, Sweden
}

\offprints{vallery.stanishev@ist.utl.pt} 

\authorrunning{Stanishev et al.}

\date{Received ;accepted}

\abstract
{Massive galaxy clusters at intermediate redshift can magnify the flux
  of distant background sources by several magnitudes.  } 
{ We exploit  this effect to search for lensed distant supernovae that may 
  otherwise be too faint to be detected.}
{A supernova search was conducted at near infrared wavelengths using
the ISAAC instrument at the VLT.  The massive galaxy clusters
\object{Abell 1689}, \object{Abell 1835}, and \object{AC114} were
observed for a total of 20 hours to search for supernovae in
gravitationally magnified background galaxies. The observations were
split into individual epochs of 2 hours of exposure time, separated by
approximately one month. Image-subtraction techniques were used to
search for transient objects with light curve properties consistent
with supernovae, both in our new and archival ISAAC/VLT data.  The
limiting magnitude of the individual epochs was estimated by adding
artificial stars to the subtracted images. Most of the epochs reach
90\% detection efficiency at $SZ(J)\simeq 23.8-24.0$ mag (Vega).  }
{Two transient objects, both in archival images of
\object{Abell 1689} and \object{AC114}, were detected. The transient in
\object{AC114} coincides~-- within the position uncertainty~-- with an X-ray
source and is likely to be a variable AGN at the cluster redshift. The
transient in \object{Abell 1689} was found at $SZ=23.24$ mag,
$\sim0.5$\arcsec away from a galaxy with photometric redshift $z_{\rm
gal}=0.6 \pm 0.15$. The light curves and the colors of the transient
are consistent with a reddened Type IIP supernova at redshift
$z=0.59\,\pm 0.05$.  The lensing model of \object{Abell 1689}
predicts $\sim$1.4 mag of magnification at the position of the transient,
making it the most magnified supernova ever found and only the second
supernova found behind a galaxy cluster.}
{Our pilot survey has demonstrated the feasibility to find distant
  gravitationally magnified supernovae behind massive galaxy
  clusters. One likely supernova was found behind \object{Abell 1689},
  in accordance with the expectations for this survey, as shown in an
  accompanying analysis paper.  
}


\maketitle

\section{Introduction}

Gravitational lensing by massive galaxy clusters provides unique 
opportunities to study extremely distant background objects. 
The modeling of the most massive galaxy clusters  at intermediate redshifts
shows that the flux of background sources with lines-of-sight
close to the critical lines can be magnified by more than 3 magnitudes.
These powerful natural gravitational telescopes have been successfully
used to identify and study gravitationally-lensed galaxies at very
high redshift
\citep[e.g.,][]{franx97,ellis01,hu02,kneib04,richard06,richard08}.

Repeated observations of massive galaxy clusters may also allow discovery
of very distant supernovae
(SNe) at redshifts beyond $z\simeq1.5-2$, which may be too faint to be 
detected without the lensing  magnification 
\citep[e.g.,][]{kovner88,kolatt98,saini00,sullivan00,galyam02,gg03}.  \cite{galyam02}
have searched repeated observations of galaxy clusters from the $HST$
archive and found several SNe, but only one of them in a background
galaxy, which has a redshift close to that of the lensing cluster.  

Detecting and studying SNe at very high redshifts is important for several
reasons:

1. The progenitors of core-collapse (CC) SNe are massive short-lived stars. The
CC SN rate thus reflects the ongoing star formation rate (SFR) and can be used 
to measure it.
The first results on the SFR derived from CC SNe \citep{dahlen04} show
an increase in the range $z\sim0.3-0.7$, consistent with
the measurements from galaxy luminosity densities.
The magnification provided by galaxy clusters may help to detect  a population of 
 CC SNe at $z>1$ for the first time  and to obtain an independent
measurement of the cosmic SFR in a redshift range where current
theoretical predictions do not give consistent results
\citep[e.g.,][]{dahlen04,mannucci07,koba08}. It may also be possible
to put constraints on the rates of the very luminous Type IIn and
pair-production SNe, and on Type Ic hypernovae.  In addition, there is growing evidence
that Type IIP SNe could be calibrated as standard candles
\citep{hamuy02,nugent06,poznanski08,olivares08} and may turn out to be more useful 
cosmological probes than Type Ia SNe (SNe Ia) at high redshift because
SNe Ia may be very rare or even absent at $z>2-3$ \citep{koba98}.  

2. Type Ia SNe have been used to measure the cosmological parameters;
however,  the scenario(s) leading to the explosion of SNe Ia is still unknown.  Models predict that
different progenitor scenarios should have different delay times
$\tau$ between the onset of star formation and the SN~Ia explosion. By measuring
the SN~Ia rate and comparing with the SFR, it is possible to set
constraints on $\tau$, hence on progenitor scenarios \citep[see
  e.g.,][]{strolger04}. The predicted SN~Ia rate is most sensitive to
the delay time for $z> 1.5$, but, as for CC SNe, different SN~Ia rate
predictions significantly diverge at $z> 1.5$
\citep{koba98,dahlen08,mannucci07,koba08,neill06,scan05}.
\cite{strolger04} and \cite{dahlen08} find a long delay time $\tau\sim3-4$~Gyr
from the low number of SNe found at $z>1.4$. This implies that SN~Ia
progenitors are old, low-mass stars and that there should be a steep
decline in the SN~Ia rate at $z>1.5$ (as in the model of
\citealt{koba98}).  However, \cite{mannucci05} and \cite{sullivan06}
present evidence for a dependence of  the SN~Ia rate on the SFR, thus
relating a significant fraction of SNe Ia to young massive stellar populations.
It is expected then that the SN~Ia rate will increase at $z>1.5$,
although \cite{mannucci07} suggest that a considerable fraction of SNe
Ia may be missed because of dust extinction.  Measurement of the SN~Ia rate at
$z>1.5$ will thus be critical for distinguishing between the models.

3. Modeling of the mass distribution of the best-studied clusters
shows that the magnification of lensed SNe can be estimated to an
accuracy of $\leq0.25$ mag  or better depending on the SN location and the 
accuracy of the mass model \citep{kneib96,saini00}.  Thus, lensed SNe
can be used as distance indicators and significantly increase the
leverage of the current SN~Ia Hubble diagram.

4. The time delay between multiple images of highly magnified galaxies could be as
shotr as weeks or months. Detection of multiple SN images is then possible, and
the time delay between them could be used to
constrain the Hubble parameter \citep{refsdal94}. Because of the transient 
 nature of SNe, the uncertainty in the time delay will
be small \citep{goobar02,morstell05}.

In this paper we describe a pilot near-infrared (NIR) survey designed
to find gravitationally magnified SNe in galaxies behind the
massive, well-studied galaxy clusters \object{Abell 1689},
\object{Abell 1835} and \object{AC114}. Here, we describe the
observing strategy, data reduction, and the SN search efficiency of the
survey. In an accompanying paper \citep[][hereafter Paper
II]{goobar08a} we present a detailed analysis of the survey and the
implications for future deeper and wider surveys. A feasibility study
of the potential for improving the mass models of clusters of galaxies
and measuring cosmological parameters using lensed SNe will be
presented in Riehm et al. (in preparation, Paper III).
 Throughout the paper we assume the concordance cosmological
model with $\Omega_M=0.3$, $\Omega_\Lambda=0.7$, $w=-1$ and $h=0.7$,
and we use magnitudes in the Vega system.

\section{Observations}

For this pilot survey, we selected three of the best-studied
massive galaxy clusters at intermediate redshift: \object{Abell 1689}
($z=0.183$), \object{Abell 1835} ($z=0.253$), and \object{AC114}, also
known as \object{Abell~S1077} ($z=0.312$).  The mass distributions of these
clusters have been  extensively modeled
\citep{broadhurst05,richard06,limousin07} and the magnification maps
can be  computed as a function of source redshift.  Deep multi-color optical and NIR
observations are available for all clusters, including $HST$ data,
making it possible to compute photometric redshifts of most of the galaxies in
the FOV. In addition, spectroscopic redshifts of multiple
images of many strongly lensed background galaxies are available.

At the targeted redshifts, beyond $z\sim1$, the UV/optical wavelengths,
where most of the SN luminosity is emitted are redshifted to the near
infrared (NIR) domain, and the search should be ideally done in the
NIR. Even when including the lensing magnification, supernovae at $z\sim1$
are expected to be fainter than 23 mag in the NIR, and thus 8-m class
telescope observations are needed. For the observations described in
this paper, we used VLT/ISAAC \citep{moorwood98}.
The Einstein radii of the three clusters
at high redshift are $\sim30-50$\arcsec,
and the large-magnification regions fit well 
into the ISAAC 2.5\arcmin$\times$2.5\arcmin\ field-of-view (FOV).
In addition, archival NIR observations at VLT/ISAAC under excellent seeing 
conditions are available for all clusters and can be used as reference 
images for the SN search.

To test the feasibility of the project, we used archival VLT/ISAAC
observations of \object{Abell 1835} obtained with an $SZ$
filter\footnote{The $SZ$ filter is a broadband filter with central wavelength
1.06$\mu$m and width 0.13$\mu$m. It is most often used as an 
order-sorting filter for ISAAC NIR spectroscopy but is also offered as an
imaging filter.}  \citep{richard06}.  The observations  were
obtained in two runs one month apart and with total exposure times of
$\sim2$ and $\sim4$ hours.  The two combined images were subtracted
and searched for transient sources.  None were found, but these
observations allowed us to test the data reduction and the image
subtraction scheme on real data, and to realistically estimate the SN
survey depth. By adding artificial stars we could establish that we
would have a good detection efficiency down to $SZ\simeq24.5$ mag.
Using
$SZ\simeq24.5$ mag as the projected photometric depth of the survey, the
cluster magnification map from \cite{richard06} and the absolute
magnitudes of the different SN types, a first estimate of the number
of SNe that could be detected in our survey was obtained following the
procedure in \citetalias{goobar08a}.  The results were encouraging and
we estimated that a few SNe per cluster per year could be
detected. For the highest redshifts, besides Type Ia SNe, the survey
 should be most sensitive to Type IIn and IIL SNe, which have very blue
spectral energy distribution (SED) around maximum light.
Close to peak brightness, Type IIP SNe should also be within
reach. Since significant star formation may take place in the blue
arcs of the strongly lensed galaxies behind the clusters, several
magnified Type II SNe are expected in our survey, possibly with
multiple images within our observational window.

After the preliminary study, we decided to observe our targets with 2
hours of total exposure time per observation with 30 days
cadence  over the period when the clusters were observable. 
The search was thus performed in a `rolling search' scheme, 
where new SNe are discovered and old SNe are followed using 
the same observations. \object{Abell 1689} and
\object{Abell 1835} were observed with the $SZ$ filter and \object{AC114} with $J$  
to be able to use the archival VLT/ISAAC observations as references.
Details of our observations are given in Table~\ref{t:detlim}
(the 2007 data), along with other NIR and optical data from the ESO
archive. No spectral or multi-color observations were obtained and for this
pilot run, the SN identification is done using the information on
light curve shapes, absolute magnitudes and host galaxy type and
redshift, if known.

\section{Data reduction}

Because of the presence of bright galaxies with extended halos at 
the cluster centers, the best observing strategy is beam-switching.
However, this observing mode doubles the execution time needed for the observations.
We therefore used a simple dithering scheme without beam-switching for
separate sky measurements. The exposures were quasi-randomly dithered
within a box of size 40\arcsec -- larger than the typical size of
the bright cluster galaxies -- and we adopt a slightly revised version the reduction
procedure described by \cite{richard06} in order to optimize the
background subtraction and the detectability of faint point sources.

All images were first corrected for the `electrical ghost' effect
with the recipe {\it ghost} within the ESO software library {\it
Eclipse}\footnote{available at {\tt
http://www.eso.org/projects/aot/eclipse/}}. The effect consists of an
additional signal, which on a given row is proportional to the sum of
the intensity along this row and the row 512 rows away. It has a
reproducible behavior and can be removed with a simple
algorithm\footnote{see {\tt http://www.eso.org/sci/facilities/paranal/
\\ instruments/isaac/doc/drg/html/node33.html}}.  Dark current and
flat-field corrections were applied using observations obtained as a
part of the ISAAC standard calibration plan.  The most challenging
part of the NIR data reduction is the background subtraction. We
performed this with the XDIMSUM package in IRAF\footnote{All data
reduction and calibration was done in IRAF and with our own programs
written in IDL. IRAF is distributed by the National Optical Astronomy
Observatories, which are operated by the Association of Universities
for Research in Astronomy, Inc., under cooperative agreement with the
National Science Foundation.} in a 3-step process:

\begin{table}[t]
\centering 
\caption[]{Data used to search for transient objects.}
\begin{tabular}{lccc}
\hline
\hline
\noalign{\smallskip}
 Date      & Exposure & Seeing & 90\% detection  \\
          & [min] & [arcsec] & efficiency [mag]    \\
\hline
\noalign{\smallskip}
\multicolumn{4}{c}{Abell 1689 -- VLT/ISAAC $SZ$-band}\\
\hline
\multicolumn{4}{c}{pointing 1}\\
2003 02 09$^a$  &   159 &   0.52     & 24.28, transient \\
2003 04 27      &    43 &   0.43     & 24.28, transient  \\
2004 01 13      &    43 &   0.52     & 23.58, non-detect   \\
2004 02 14      &    43 &   0.58     & 23.64, non-detect \\
\hline
\multicolumn{4}{c}{pointing 2}\\
2003 01 16      &    43 &   0.58     & 23.48\\
2003 02 15      &    43 &   0.50     & 23.60 \\
2003 04 27      &    86 &   0.44     & 23.64  \\
2004 01 12      &    43 &   0.55     & 23.64\\
\hline
\multicolumn{4}{c}{pointing to cluster core}\\
2007 04 08      &   117 &   0.64 &  23.95  \\
2007 05 14/15   &   117 &   0.65 &  23.95  \\
2007 06 06      &    39 &   0.70 &  23.15  \\
\hline
\multicolumn{4}{c}{AC 114 -- VLT/ISAAC $J$-band}\\
\hline
2002 08 20 &      108   &   0.49 & 23.87 \\
2007 07 13$^b$ &  234   &   0.43 & 24.04 \\
2007 08 09 &      117   &   0.73 & 23.79 \\
2007 09 02 &      117   &   0.55 & 23.83 \\
2007 09 28 &      117   &   0.46 & 24.04 \\
\hline
\multicolumn{4}{c}{Abell 1835 -- VLT/ISAAC $SZ$-band}\\
\hline
area 1          &       &        &         \\
2004 04 20      & 231 	&   0.49 & 24.06   \\ 
2004 05 15	& 135 	&   0.62 & 24.06   \\
2007 04 18	& 117 	&   0.79 & 23.80   \\ 
2007 05 18	& 78  	&   0.74 & 23.83   \\
2007 07 18	& 117 	&   0.62 & 23.80   \\
area 2    	&     	& 	 &	   \\
2007 04 18      & 117 	&   0.79 & 23.70   \\
2007 05 14/18   & 60  	&   0.80 & 23.45   \\
2007 07 18      & 117 	&   0.62 & 23.70   \\
\hline

\end{tabular} \\
$^a$ -- average of observations obtained on 5,11 and 15 February;\\
$^b$ -- average of observations obtained on 11,12,13 and 15 July.
\label{t:detlim}
\end{table}

\begin{enumerate}

\item For each image, the 8 exposures closest in time were scaled to
  the same mode and median combined to estimate the sky
  background. The combined image was scaled to the mode of the image
  from which the background is to be removed and was subtracted from
  it. The sky-subtracted images were then combined 
with integer pixel shifts to preserve the
  characteristics of the noise. The ISAAC pedestal is highly
  variable in time and is very difficult to remove accurately. In some
  of the images, significant residuals were found along the bottom and
  the middle rows of the array, where the readout of the two halves of
  the array starts. It was then necessary to correct for this effect
  before the final combination. At the first step this was done by
  simply taking a median along the image rows with $2\sigma$ clipping
  and subtracting it.

\item The combined image from Step 1 was used to detect objects
  and create an object mask. For this XDIMSUM uses a simple method
  where all pixels above a given threshold, typically 2 times 
  the standard deviation of the sky, are
  assumed to belong to objects. The mask is then deregistered to each
  individual image. The same operations as in Step 1 were then
  repeated, but now all pixels belonging to objects were excluded when
  performing the background estimation and residual pedestal
  subtraction. The only difference was that, instead of the median,
  the mode along the image rows was used to remove the residual pedestal.

\item The combined image from Step 2 was used to create an improved
  object mask using the more sophisticated OBJMASK task. This task
  detects connected pixels and can also apply object growing, which in
  our case was important for masking the extended halos of the
  cluster galaxies. The background subtraction was repeated with the
  improved object mask to arrive at our final sky-subtracted images.

\end{enumerate}

For the final combination, we assigned weights to the individual images
in order to optimize the detection of faint point sources. The weights
are inversely proportional to the product of the square of the seeing,
the variance of the sky background and the multiplicative factor that
brings the images to the same flux scale. The archival ISAAC observations, which we used as
reference images and when possible to search for SNe, 
were reduced following the same recipe.

The photometric calibration of the images was obtained with standard
stars from \cite{persson98} and a few from \cite{hawarden01}
observed in each photometric night as part of the ISAAC standard
calibration plan. Because the standard stars do not have calibrated $SZ$
magnitudes, we used $J$-band magnitudes to calibrate the
$SZ$ observations of \object{Abell 1835} and \object{Abell 1689}.
 The majority of the \cite{persson98} stars have $J-H\simeq0.3$
mag, which corresponds to late F to early G type stars. Synthetic $SZ$
and $J$ photometry was computed using spetrophotometry of F8V to G5V
stars from \cite{pickles}, a synthetic spectrum of the Sun from Kurucz
web site\footnote{available at {\tt http://kurucz.harvard.edu}} and
the Vega spectrum from \cite{bohlin07}, assuming all colors of Vega to
be zero. The computed synthetic $SZ-J$ colors ranged between 0.1 and
0.2 mag, and we  corrected the zero points estimated from the
$J$ magnitudes of the \cite{persson98} stars by $+0.15$ mag to derive
the $SZ$ zero points.  
A correction of the same order was
also estimated based on synthetic photometry of elliptical galaxies,
as described in \cite{richard06}.

\section{The transient search and survey depth}

 \begin{figure}[!t]
\centering
\includegraphics*[width=8.8cm]{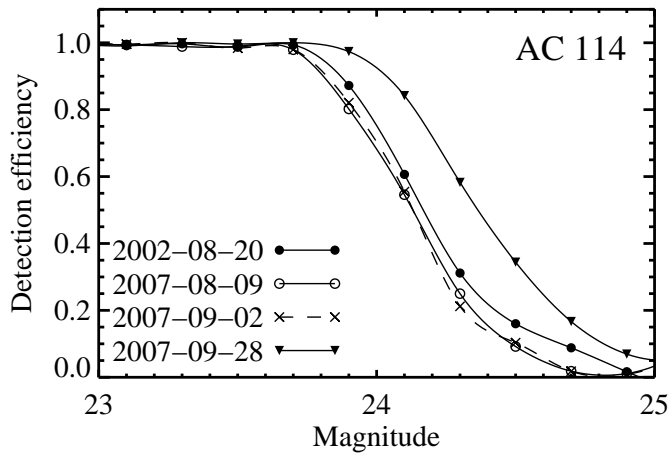}
\caption{The detection efficiency as a function of magnitude for the observations of 
\object{AC114}. The results for \object{Abell 1835} and \object{Abell 1689} 
are qualitatively the same and are not shown.}
\label{f:det_eff}
\end{figure}

To search for transient objects, the reference images were subtracted
from our new images.  The subtraction was done with Alard's optimal
image subtraction software \citep{optsub1,optsub2}, slightly modified
and kindly made available to us by B. Schmidt. The reference image is
geometrically aligned to the new image, convolved by a kernel such
that the two images have the same PSF, and then scaled and
subtracted. The residual images were searched for transient sources
with Sextractor v2.5 \citep{bertin96}.  The images were first
convolved with a Gaussian filter with FWHM equal to the seeing and
then searched for sources consisting of at least four connected pixels
$1\sigma$ above the background. 

Because we expect most SNe to be very faint and close to the detection
limit of the survey, we required the object to
be detected in at least two epochs for a secure detection.
Since SNe could have been recorded on the
archival data that were used as reference, we also searched for
negative peaks.  Part of the archival observations were obtained
during many separate runs and were also searched for transients.

The detected sources were carefully examined visually and obvious
artifacts and clear spurious detections near the underexposed image edges
 discarded. The cores of the compact, bright cluster galaxies
could often not be subtracted cleanly, so significant residuals were
left.  Therefore, all ``detections" close to badly subtracted cluster
galaxies were rejected.  After these cuts, only a few detections per
image were typically left for closer examination.  None of these
potential transients was detected in more than one epoch. The careful
visual examination (including blinking of the subtracted and the
original images) showed that all but two candidates could not be
visually distinguished from the background noise. 
That none of the other potential candidates were picked up by
Sextractor in another epoch makes us conclude that there was no
evidence that any of these were real transient detections. The
remaining two detections will be discussed in detail in the next
section.

To estimate the detection limits of the survey, for each epoch the PSF
was computed using stars in the fields and 50 artificial stars with
magnitudes between 22 and 25 were randomly added to the subtracted
images. For each epoch, this was repeated 200 times (10000 artificial
stars), and the artificial stars were recovered with the same
Sextractor parameters used to search for transients. The detection
efficiency was then computed in 0.2 mag bins as the fraction between
the number of recovered stars and the number of added stars. Because
we expect to find SNe on faint galaxies (if visible at all) with
surface brightnesses close to the background,
we only considered the artificial stars away from bright objects in
the field.
In addition, because we expect to find rather faint SNe close to the
detection limit, we only consider the central parts of the images that
received at least 90\% of the total exposure time. In
Table\,\ref{t:detlim}, we show the 90\% detection limits and in
Fig.~\ref{f:det_eff}, the detection efficiency curves for
\object{AC114} are shown as an example.

We note several complications.  For \object{AC114} our first
epoch had better seeing and photometric depth than the archival observations. 
We therefore used our first epoch as the main reference image for the search. A Type IIP SN 
that had nearly constant brightness during our observational window could still be
detected as a negative source in the subtraction of the archival observations.

\begin{figure}[!t]
\centering
\includegraphics*[width=8.8cm]{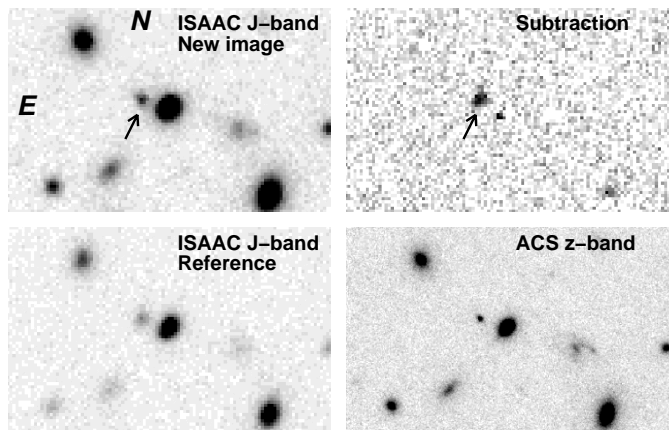}
\caption{The subtraction of the $J$-band images of \object{AC114}. The 
arrow shows the transient. The ACS $z$-band image is also shown.}
\label{f:ac114}
\end{figure}

The archival observations of \object{Abell 1689} have been obtained at
4 pointings in order to cover the whole cluster. Each pointing covers
$\sim1/4$ of our new images and so each quarter was then treated
separately using the respective reference image.  The four combined
reference images have nearly the same seeing, and the detection
efficiencies were nearly the same. In Table\,\ref{t:detlim} we give
the average detection limits.

All observations of \object{Abell 1835} should have had the same
pointing as the archival data, namely 1\arcmin\ off the cluster center
 to avoid a very bright star falling into the FOV.  However, 
a coordinate error meant that the observations were obtained at
two pointings: the intended one and the one without the offset. 
The parts of the images that overlapped with the archival data
were analyzed using the reference image from the archival data. The
detection limits are given as ``area 1" in Table\,\ref{t:detlim}. Those
parts of the images that did not overlap with the archival data were
analyzed using our observation on 18 July 2003 as the reference. This is
``area 2" in Table\,\ref{t:detlim}. Because of the different
photometric depth of the reference images, we obtain slightly
shallower detection limits for area 2.

\section{Results}

The search yielded the detection of two transient objects, one in each of the fields
of \object{Abell 1689} and \object{AC114}. Both objects were detected 
in the archival images and not in our new survey images. 

\begin{figure*}[!t]
\centering
\includegraphics*[width=18cm]{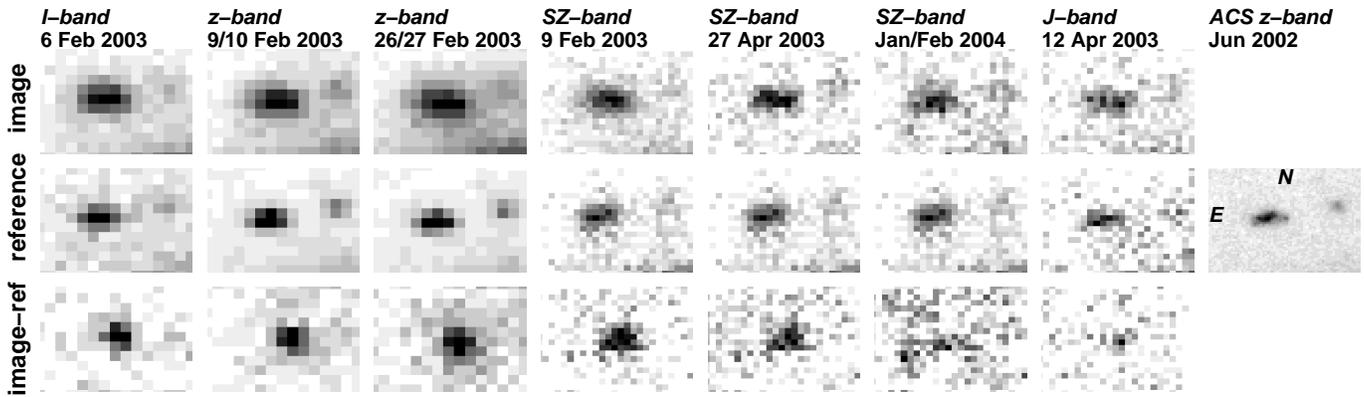}
\caption{The subtraction of the images for \object{Abell 1689}. 
 The ACS $z$-band image is also shown.}
\label{f:a1689}
\end{figure*}

\subsection{The \object{AC114} transient}

Stamps from the $J$ and $ACS$ $F850LP$-band (effectively corresponding
to Gunn $z$-band) images of the transient in \object{AC114} are shown
in Fig.\,\ref{f:ac114}. The transient is seen in all archival $J$, $H$,
and $K$ images as a compact unresolved source. It is also present --
barely resolved from the nearby galaxy -- in the ground $B$ and $V$
band images, as well as on the $ACS$ $z$-band images taken 14
November 2004. The coordinate matching of the $J$-band to the $ACS$
image shows that the variable object coincides exactly with a compact
barely resolved galaxy. It is also within the position errors of the
X-ray source \object{AC114-2} detected by \cite{martini06} in
$Chandra$ images. The spectrum of \cite{martini06} is relatively blue,
which together with the detected
X-ray emission suggest that the object is a variable
AGN. \cite{martini06} measure a redshift of $\sim0.32$, which implies
that the object is in \object{AC114}. We also subtracted the 2 epochs
of archival ISAAC $H$-band imaging obtained on 20 August 2002 (coeval
with the $J$-band observations) and 17 May 2004. The subtraction does
not show a significant flux change. AGNs may vary slowly on time scales
of years so the $H$-band results do not necessarily contradict 
the AGN hypothesis. Given all the available evidence, a variable AGN
seems the most likely explanation for the transient 
in \object{AC114}.

\begin{table}[t]
\centering 
\caption[]{Measured magnitudes of the transient in \object{Abell 1689}.}
\begin{tabular}{lccc}
\hline
\hline
\noalign{\smallskip}
 Date      & Magnitude & Filter & Telescope/ \\
           &   [Vega]        &  &  Instrument     \\

\hline
\hline
2003 02 06    &   24.09  $\pm$  0.20  & $I$  & VLT/FOSR2 \\
2003-02-9/10  &   23.93  $\pm$  0.08  & $z$  & VLT/FOSR2  \\
2003-02-26/27 &   23.94  $\pm$  0.09  & $z$  & VLT/FOSR2  \\
2003 02 09    &   23.24  $\pm$  0.08  & $SZ$ & VLT/ISAAC\\
2003 04 27    &   23.73  $\pm$  0.16  & $SZ$ & VLT/ISAAC\\
2004 01/02    &   $>24.3-24.4$        & $SZ$ & VLT/ISAAC\\
2003 04 12    &   23.61  $\pm$  0.15  & $J$  & VLT/ISAAC\\
\hline 

\end{tabular} \\
\label{t:phot}
\end{table}

\subsection{The \object{Abell 1689} transient}

The transient in \object{Abell 1689} was detected as a negative source
in the $SZ$-band subtractions that used a stack
of all archival data as a reference image.  The archival $SZ$ data were 
obtained in 4 runs
between February 2003 and February 2004 (data under ``pointing 1" in
Table \ref{t:detlim}), which made it possible to build a light curve. The
magnitude of the transient was $SZ=23.24\,\pm0.08$ on 9 February 2003. It was
half a magnitude fainter on 27 April 2003 and was not detected in the
combined January/February 2004 image. The 90\% detection limit of the
latter image is $\sim24.15$ mag. However, the images with added
artificial stars were examined and it was noticed that if the object
location is known, objects down to 24.3--24.4 mag could be detected
{\it visually}.

\object{Abell 1689} has been observed on many occasions at the VLT, including
times around the epoch at which the transient was detected. The
cluster has also been observed with WFPC2 and ACS on $HST$. 
 All these archival data made it possible to obtain four additional 
measurements of the magnitude of the transient in three bands -- 
Bessell $I$, Gunn $z$, and the NIR $J$ band.
The instrumental magnitudes of the transient were measured on reference-subtracted
images with the PSF fitting technique using DAOPHOT in IRAF. The
uncertainties were estimated by adding artificial stars with the same
magnitude as the transient and measuring their magnitudes in exactly
the same way as the transient. This procedure provides a more reliable
error estimate than the formal output of DAOPHOT because the noise in
the reference-subtracted images is correlated. As a by-product, we also
obtain the accuracy with which the positions of artificial stars could be
recovered, which we take as a measurement of the accuracy
of the measured position of the
transient.  The magnitudes of the transient are listed in
Table\,\ref{t:phot},  and Table\,\ref{t:a1689} lists details of the
observations used. Stamps of the images where the transient was
present, the references, and the difference between them are shown in
Fig.\,\ref{f:a1689}. Also shown are the non-detection in the $SZ$-band
observations during January/February 2004 and the stamp from the ACS
$z$-band image.

 The FORS2 $I$-band observations were obtained in February, June, and
July 2003 and in 2001. The deep April 2001 images served as a reference. 
 The magnitude of the transient could only be measured in the February 2003
data and the rest were too shallow for the transient to be detected.  
The FORS2 Gunn $z$ observations
were taken on 9, 10, 26, and 27 February 2003, but to increase the S/N, the  images for 
9, 10  and 26, 27 were combined.
The $ACS$ $z$-band image (obtained on 16 June 2002) was used as a
reference and yielded very clean subtractions.
For both epochs we measured the transient at  about the same brightness of $z\simeq23.9$ mag. 
The constant brightness of the transient could be further 
verified by subtracting the two FORS2 $z$-images, which left no noticeable residuals. 

The photometric calibration of the FORS2  $I$-band data was obtained
with \cite{landolt92} standards observed during the same night. The night 
was photometric, but as a consistency check, we analyzed large number of additional
FORS1/2 $I$-band observations of \object{Abell 1689} obtained during 2001 and 2003,
most of which were acquisition images for spectroscopic observations. Most of the nights
were photometric, and \cite{landolt92} standards were also observed. We used them
to calibrate a number of faint stars in the field assuming 0.04 mag per airmass 
atmospheric extinction in $I$. The typical RMS of the calibrated stars
was $\sim0.02$ mag, confirming that the nights,  including 6 February 2003, were photometric.

The calibration of the FORS2 Gunn $z$ observations was obtained 
through the primary white dwarf standard star \object{GD\,153} observed on the photometric night of 10 February 2003, 
immediately after the observations of \object{Abell 1689}\footnote{Stars with calibrated 
$z$ magnitudes from \cite{hamuy01_99em} were also observed but were all saturated.}. We used the
synthetic $VRI$ magnitudes of \object{GD\,153} from \cite{holberg06}, together with the 
relation between the Landolt $VRI$ and the Gunn $z$ magnitudes derived by \cite{kri_ir_temp}, 
to compute the $z$ band magnitude of \object{GD\,153}, finding $z=13.77\,\pm0.04$. Most of the
uncertainty comes from the RMS around the relation of \cite{kri_ir_temp}.  
We then computed the $z$ band magnitudes of the stars that were used to 
cross-check the $I$ band photometry. We assumed 0.04 magnitudes of extinction
per airmass for the z-band. These stars were then used to calibrate the
photometry of the transient in the combined 9-10 and 26-27 February 2003 images.

The single $J$-band observation was obtained on
12 April 2004. As a reference, we used a $J$-band image
from VLT/HAWK-I that was obtained as part of our new program to
search for SNe behind \object{Abell 1689} (a sequel to the pilot ISAAC
search we present here).
ISAAC $H$-band data from April and May 2003 were also available, 
but were too shallow to detect the transient.

The coordinates of the transient were inferred from 
astrometric solution using coordinates of the field objects from the
{\it Sloan Digital Sky Survey} Data Release 6. The average of the six
measured positions is $\alpha_{2000} = 13^\mathrm{h}11^{\rm
  m}$28\fs170 ($\pm$0.002) and $\delta_{2000} =
-01$\degr19\arcmin25\farcs128 ($\pm$0.012), where the uncertainties
are the standard deviation of the mean.

\section{Discussion}

\begin{table}[!t]
\centering 
\caption[]{Additional archival observations used to measure the
 brightness of the \object{Abell 1689} transient, with 
HST/ACS data and those used as reference images not listed.}
\begin{tabular}{lccc}
\hline
\hline
\noalign{\smallskip}
 Date      & Exposure & Seeing & Note \\
           & [min] & [arcsec] &     \\

\hline
\noalign{\smallskip}
\multicolumn{4}{c}{VLT/ISAAC $SZ$-band}\\
\hline
2003 02 09$^a$  &   159 &   0.52     & transient \\
2003 04 27      &    43 &   0.43     & transient  \\
2004 01/02      &    86 &   0.55     & non-detection  \\
\hline
\multicolumn{4}{c}{VLT/ISAAC $J$-band}\\
\hline
2003 04 12  &   132 &   0.47     & transient  \\
\hline
\multicolumn{4}{c}{VLT/FORS2 $I$-band}\\
\hline
2003 02 06  &   36  &   0.66     & transient  \\
\hline
\multicolumn{4}{c}{VLT/FORS2 $z$-band}\\
\hline
2003 02 09/10  &   256  &   0.60 & transient  \\
2003 02 26/27  &   324  &   0.70 & transient  \\
\hline
\end{tabular} \\
$^a$ -- average of observations obtained on 5,11 and 15 February.
\label{t:a1689}
\end{table}

 The search yielded two transient objects, one in the field of \object{AC114} and one in 
\object{Abell 1689}.  The transient in \object{AC114} is very likely a
variable AGN at the cluster redshift. Below we argue that the 
transient in \object{Abell 1689} was probably a Type IIP SN at 
redshift $z\simeq0.6$ and that other alternatives, including a 
supernova at the cluster redshift, are unlikely.

\begin{figure}[!t]
\centering
\includegraphics*[width=8.8cm]{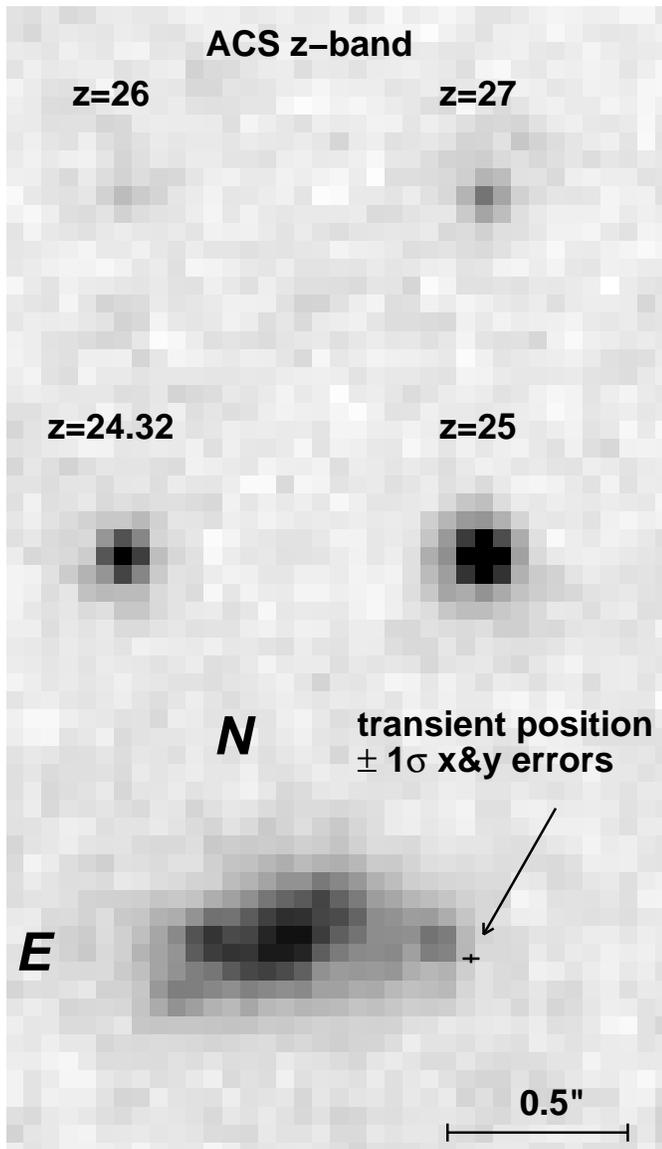}
\caption{ACS $z$-band image of the putative host galaxy. The cross marks the 
average position of the transient from the six detections and the size along the $x$ and $y$ 
axes is the standard deviation of the mean. Four artificial stars with the indicated magnitudes were added to the image for comparison.}
\label{f:tranpos}
\end{figure}

 Figure\,\ref{f:tranpos} shows a zoom of the deep $ACS$ $z$-band image
(16600~s total exposure time; the other coeval deep $ACS$ F475W, F625W
and F775W images show the same galaxy morphology) 
centered on the region near the position of
the transient behind \object{Abell 1689}. Four artificial
stars with magnitudes 24.32 (the transient magnitude on the ground
based $z$ images), 25, 26, and 27 mag were also added to demonstrate
the image depth.  The uncertainty of the position of the transient, 
the standard deviation of the mean, was estimated from 
the coordinate transformations between the images where the 
transient was detected and the $ACS$ image. 
This uncertainty
  is $\sim0.5$ $ACS$ pixels (1 $ACS$ pixel = 0.05\arcsec), which is consistent
  with the uncertainties of the individual position measurements
  $\sim1-1.5$ $ACS$ pixels as
computed by adding in quadrature the uncertainty of the positions 
of the transient as estimated from the recovered
artificial stars, and the uncertainty of the coordinate transformations
to the $ACS$ image.
  
 The transient was not detected in the deep June 2002 $ACS$ images
to a limiting magnitude $z\simeq26.5$, but a long blazar outbursts
with an amplitude $\geq2.5$ mag  could, in principle, offer an
alternative explanation of the transient. The synthetic $I-SZ$,
$z-SZ$, and $SZ-J$ color indices of the QSO template of
\cite{glikman06} match those of the transient, if the template is
moderately reddened with $E(B-V)\simeq0.2$ and redshifted to
$z\sim0.6-0.7$.  Some blazars do show repeated large-amplitude
outbursts on a time scale of months to a few years, but most show
variability with amplitude $<1$ mag.  However, the archival
$HST/WFPC2$ images taken in June 1996, June 1997, and April 2007 (the
lattest coeval with our new ISAAC data), the deep 2001 VLT, and 2002
$HST/ACS$ observations do not show any object at the position of the
transient.  Furthermore, the closest galaxy has an irregular
shape, but the position of the transient is $\sim0.5$\arcsec away from
its center and  does not coincide with any of the outer spots.
 Along with the non-detection of any X-ray source close to
the transient position \citep{martini06}, the variable
AGN hypothesis seems quite unlilkely.

A galactic origin for the transient is also unlikely. Any plausible
candidate would be either bright enough  to be detected in the
deep ACS images (e.g. novae) or  would have faded away faster
than the transient (e.g. flaring late type stars). For example, WZ
Sge-type dwarf novae are faint in quiescence, $M_V\simeq+11$ mag
\citep{harrison04},  and may not be detected in the deep ACS image
if farther away than 12-13 kpc. \object{WZ Sge} stars show
large amplitude outbursts (up to $\sim7$ mag), which may bring them
above the detection limit.  However, the duration of the outbursts is
typically less that 80 days and the brightness decreases faster than
observed in the \object{Abell 1689} transient.  Additional
arguments against this scenario are the observed red colors of the
transient and its $I$ magnitude. Dwarf novae at outburst have blue
colors in contrast to the observed red colors of the transient. The
absolute $I$ magnitude of \object{WZ Sge} at outburst maximum is
$\sim+5$ \citep{harrison04,howell04} and the observed $I\simeq24.1$
would imply a distance of about 70\,kpc, which is unlikely.

The slow decline
of the $SZ$ light curve suggests a Type IIP supernova. The interaction
of the expanding ejecta with circumstellar material in Type IIn or
possibly in other SNe types \citep[e.g. SNe 2002ic and
2005gj,][]{hamuy03,aldering06} may also lead to slow luminosity
decline. A Type~Ia SN at a few months after maximum 
should also be considered as a possible candidate for the 
observed transient possibility.
The brightness of prompt optical emission of gamma ray bursts 
typically declines much faster than the transient in \object{Abell 1689} and  the most plausible
conclusion is that the transient was a supernova.

\begin{figure}[!t]
\centering
\includegraphics*[width=8.8cm]{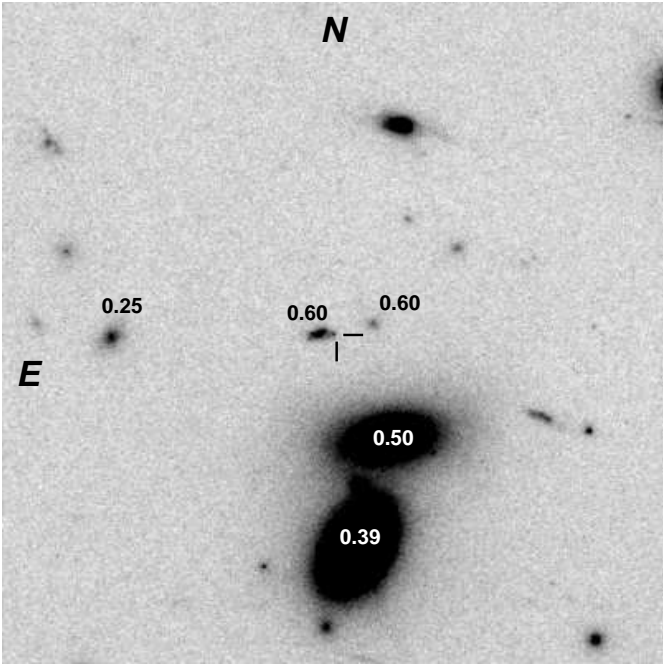}
\caption{Zoom of the ACS $z$-band image around the transient in \object{Abell 1689}. 
The image size is 20$\times$20\arcsec or 
60$\times$60\,kpc at the cluster redshift. The position of the transient is marked and 
the estimated photometric redshift of some galaxies is also shown.}
\label{f:zoom}
\end{figure}

Figure\,\ref{f:zoom}  displays a 20$\times$20\arcsec region centered on the
transient.
At the cluster redshift, this corresponds to  
60$\times$60\,kpc. None of the galaxies close to the transient have
 spectroscopic redshifts.  The photometric redshift 
 of the closest galaxy based on $B$-to-$K$-band photometry
 \citepalias[for details see][]{goobar08a} is $z=0.60\,\pm0.15$. The
 best-fitting template is based on single starburst evolved by 50~Myr
 and only slightly reddened. The other possible host galaxies, the
 faint objects right of the transient, and the two bright ellipticals
 have photometric redshifts between 0.4 and 0.6. Therefore
 core-collapse SN in a cluster or foreground galaxy is unlikely.
 Intercluster, hostless SN~Ia at about 80-100 days past maximum light
 could match the observed brightness of the transient.  However, the
 late-time ($>150$ days past maximum) brightness of SNe Ia typically
 declines by $\sim1.5$ mag per 100 days and even faster between 70 and
 150 days. Thus the transient should have declined by about 1 mag over
 the 77 days between the two $SZ$ observations, which is twice more
 than observed.  Thus, we find it quite unlikely that the  
transient could be a supernova
 at the cluster redshift.

To further investigate the nature of the transient we fitted its
measured magnitudes with several templates based on the well-observed
Type IIP SNe \object{2004et} \citep{sahu06}, \object{1999em}
\citep{hamuy01_99em,kris09}, \object{2003hn} \citep{kris09}, and
\object{2001cy} \citep{poznanski08}. The SNe were carefully selected
to sample all light curve morphologies of Type IIP SNe. We also used
\object{2005gj} \citep{prieto07}, which  had a slow decline in brightness 
after maximum light because of interaction with
the circumstellar medium. The three SNe
with NIR photometry (\object{2003hn}, \object{1999em}, and
\object{2005gj}) also allowed us to test redshifts below 0.4. Because
only \object{1999em} have (few early) NIR spectra, the NIR magnitudes
of these SNe were fitted with a cubic spline and the resulting fits
were used to compute the cross-filter $K$-corrections \citep{kim96}.

\begin{figure}[!t]
\centering
\includegraphics*[width=8.8cm]{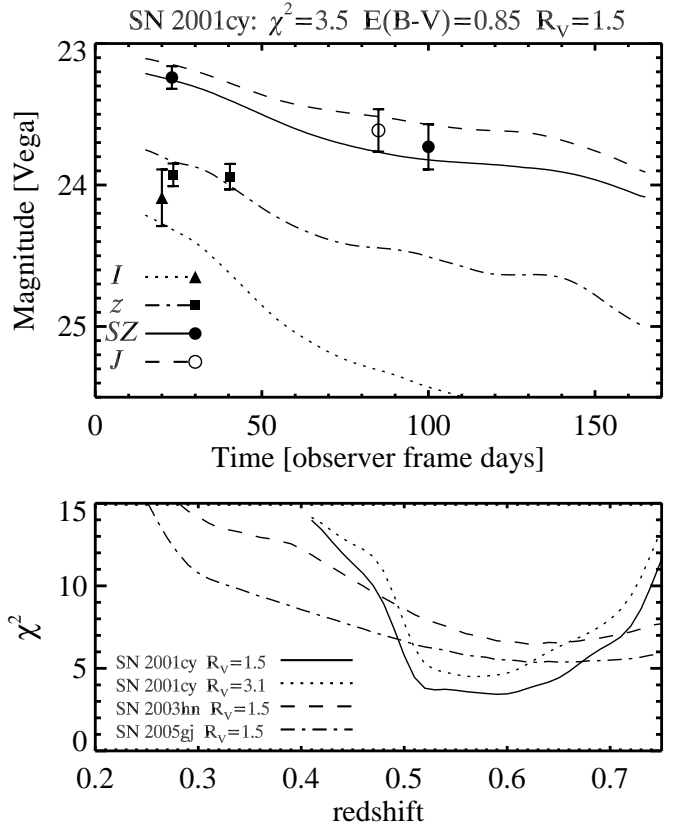}
\caption{{\bf Upper:} Best fit of the transient magnitudes with a light 
curve template based on \object{SN 2001cy}. {\bf Lower: } $\chi^2$ $vs.$ 
redshift for three core-collapse SN light curve templates.}
\label{f:fit}
\end{figure}

The templates were  used to perform lightcurve fits
 on a grid of redshifts and reddening
parameters.  Since we only have six measurements, we tried three
fixed values of $R_V$ and only allowed  $E(B-V)$ to vary, 
along with 3 other lightcurve parameters: redshift, time of maximum, and
peak magnitude.  The best
fit, shown in Fig.\,\ref{f:fit}, was obtained with the template based
on \object{SN 2001cy} redshifted to $z=0.59\,\pm0.05$ and reddened by
non-standard dust in the host galaxy with $E_{B-V}=0.85$ and
$R_V=1.5$. It should be noted that the uncertainty of the redshift
estimated from the light curve fit is the formal error from the fit
and  may be somewhat underestimated. 
 The lower panel of Fig.\,\ref{f:fit} show the 
$\chi^2$ as a function of the redshift for the three best-fitting
templates, all for $R_V=1.5$, as well as \object{SN 2001cy} for the
standard Milky Way $R_V=3.1$. It is interesting to note that the
best fits were  obtained for low $R_V=1.5$ and redshift around
0.6 and that lower redshifts  result in much poorer fits. SNe \object{1999em}
and \object{2004et} do not fit the data well because their $R$ and $I$
brightness are either constant or increases on the plateau phases,
while the transient $SZ$ magnitude decreases.

The low best-fit value of $R_V$ deserves special comment. Recent
studies of individual nearby SNe Ia and statistical analysis of large
samples of SNe Ia \citep[e.g.,][]{nancy06,nancy08,snls1,nobili08} and
now also for Type IIP SNe \citep{poznanski08,olivares08} find that
dust in SN host galaxies has $R_V$ significantly lower than the
average Milky Way value of $\sim3.1$. It is possible that the dust
that obscured the transient had properties corresponding to low $R_V$,
possibly as a result of multiple scattering by circumstellar dust
clouds \citep{goobar08}. It should be noted that in the fitting
process, we used the Milky Way extinction law parametrization by
\cite{fitzpatrick99}, which was derived using observations of
stars obscured by dust with $R_V$ between 2.3 and 5.3\footnote{The
other widely used parametrization of \cite{ext_law} was 
obtained with stars obscured by dust with $R_V$ between 2.75 and 5.3}. Any use of this
parametrization to compute the extinction laws for $R_V<2.3$ thus
involves extrapolation, which may be inaccurate, especially if $R_V\ll2.3$. 
 A more reliable approach  may be to use a physical model of the interstellar 
dust \citep{weingartner01}, but this is beyond the scope of this paper.

 To estimate the absolute magnitude of the transient, we correct the peak magnitude 
that we estimated from the light curve fit for the dust extinction and gravitational 
magnification of the cluster.  For this we used the accurate parametric mass 
distribution model of \object{Abell 1689} from \cite{limousin07}. The model is based 
on  strong lensing analysis and uses 34 multiply imaged systems, 24 of them with 
newly determined
spectroscopic redshifts.  As discussed by \cite{limousin07}, the large number 
of constraints used to derive the mass model make \object{Abell 1689} the 
most reliably reconstructed cluster
to date. For more details on the mass model of  \object{Abell 1689} see \cite{limousin07}.

The cluster magnification at the position of the transient for redshift 
$z=0.6$ is 1.36 $\pm0.03$ mag, where the uncertainty is estimated from the 
uncertainties of the mass model. For redshift $z=0.45$ and $z=0.75$, the 
magnifications are 1.13$\pm0.02$ and 1.60$\pm0.04$, respectively. The uncertainty 
from the mass model is much smaller than the
uncertainty stemming from the photometric redshift estimation and we adopt 
a magnification of 1.4$\pm0.3$ mag to estimate the absolute magnitude of the transient.
Taking  the dust extinction into account, we
find absolute $V$ magnitude of the transient $M_V\sim-17.6 \pm0.3$, which is normal for a Type
IIP SN \citep[e.g.,][]{richardson02}. 
 The transient redshift inferred from the template fit matches the photometric 
 redshift of the nearest galaxy very well, which is the most likely host of the SN. 
 With its young stellar population
 inferred from the best photometric redshift fitting galaxy 
 template and the irregular morphology, this
 galaxy is likely to produce core-collapse SNe.

\section{Conclusions}

 We have presented
the first dedicated rolling search for gravitationally magnified SNe behind
galaxy clusters.
The search was
conducted in the near infrared wavelengths using the ISAAC instrument
at the VLT.  The massive galaxy clusters \object{Abell 1689},
\object{Abell 1835} and \object{AC114} were observed for a total of 20
hours, split into individual epochs of 2 hours exposure time each
separated by approximately one month. Image subtraction was used to
search for transient objects with light curves consistent with
supernovae.  The analysis of the combined data set of new observations,
plus 20 more hours of archival data, yielded the detection of two
transient objects, both found in archival images of \object{Abell
1689} and
\object{AC114}. The transient in \object{AC114} coincides 
with an X-ray source and is probably a variable AGN at the
cluster redshift. The transient in \object{Abell 1689} was found 
with an $SZ$ magnitude of 23.24 and $\sim0.5$\arcsec away from a galaxy with photometric
redshift $z_{\rm gal}=0.6 \pm 0.15$. The light curves and the colors
of the transient are consistent with it being a reddened Type IIP
supernova at redshift $z=0.59\,\pm 0.10$.  The lensing model of
\object{Abell 1689} predicts 1.4$\pm0.3$ mag of magnification at the position
of the transient, making it the most magnified supernova ever found
and only the second supernova found behind a galaxy cluster. Without
the magnification, this SN would not have been detected  in our NIR search.

The limiting magnitude of the individual epochs was estimated by
adding artificial stars to the subtracted images. Most of the epochs
reach 90\% detection efficiency at $SZ(J)\simeq 23.8-24.0$ mag
(Vega). The detailed analysis presented in the accompanying
\citetalias{goobar08a} indicates that our survey should have detected
between 0.8 and 1.6 SNe with the peak probability corresponding to
Type IIP SNe at a redshift of $z\sim0.7$. It is encouraging that this
is what we actually found.

Our pilot survey has demonstrated the feasibility to find distant SNe
gravitationally magnified by massive galaxy clusters. The survey
reached a detection threshold of $\sim23.8-24.0$ mag. A deeper survey
of several clusters with an NIR camera with a larger field-of-view should yield an order
of magnitude increase in the number of detected SNe, a significant
fraction of which will be at redshifts $z>1.5$ \citepalias{goobar08a}.

\begin{acknowledgements} 

VS acknowledges financial support from the Funda\c{c}\~{a}o para a Ci\^{e}ncia e a Tecnologia. 
AG, VS and SN acknowledge support from
the Gustafsson foundation. KP gratefully acknowledges support from the Wenner-
Gren Foundation.  AG and EM acknowledge financial
support from the Swedish Research Council. JPK thanks CNRS
for support, as well as the French-Israeli council for Research,
Science and Technology Cooperation. We thank Dovi Poznanski for providing us the
data for \object{SN 2001cy}.

\end{acknowledgements}



\end{document}